# Competition of L2$_1$ and XA Ordering in Fe$_2$CoAl Heusler Alloy: A First-Principles Study

Aquil Ahmad$^*$, A. K. Das, and S. K. Srivastava

*Department of Physics, Indian Institute of Technology Kharagpur, India-721302*

$^*$*Email:* aquil@phy.iitkgp.ernet.in

**Abstract:** The physical properties of Fe$_2$CoAl (FCA) Heusler alloy are systematically investigated using the first-principles calculations within generalized gradient approximation (GGA) and GGA+U. The influence of atomic ordering with respect to the Wyckoff sites on the phase stability, magnetism and half metallicity in both the conventional L2$_1$ and XA phases of FCA is focused in this study. Various possible hypothetical structures viz., L2$_1$, XA-I, and XA-II are prepared by altering atomic occupancies at their Wyckoff sites. At first, we have determined the stable phase of FCA considering various non-magnetic (or paramagnetic), ferromagnetic (FM) and antiferromagnetic (AFM) configurations. Out of these, ferromagnetic (FM) XA-I structure is found to be energetically most stable. The total magnetic moments per cell are not in agreement with the Slater-Pauling (SP) rule in any phases; therefore, the half-metallicity is not observed in any configurations. However, FM ordered XA-I type FCA shows 78% spin polarization at E$_F$. Interestingly, the results of XA-I type FCA are closely matched with the experimental results.

## 1. Introduction

Traditional electronic devices, which are mostly based on semiconductors, work on the electronic charge flow and lead to the consumption of higher power. The spin degree of freedom along with the charge of electrons could be exploited as an alternative in the existing semiconductors to minimize power consumption, maximize speed of operation and add the non-volatility in the electronic devices [1, 2]. As a thumb rule, materials with high spin polarization are required in spin-based electronics called spintronics [1, 3-6]. Many full-Heusler alloys (HAs) of type X$_2$YZ (where X and Y are transition metal elements, and Z is an s-p element) exhibit half-metallicity, which means that one of the spin channels is gapless at the Fermi energy (E$_F$), while the other one possesses a semiconducting or insulating band gap. Thus, the half-metallic HAs would be 100% spin-polarized. It is widely known that the high spin polarization (P) plays a key role in performance of spintronics devices such as magnetic tunnel junction (MTJ) or spin valves [7]. The inclusion of Fe$_2$CoAl Heusler alloy may enhance the giant magnetoresistance or tunneling magnetoresistance as compared with normal ferromagnetic transition metals, where spin polarization (P) is about 30-40%. The half-metallicity is ascribed to the strong hybridization between the *d* orbitals of the two transition metals [8]. The other features of those materials are of high Curie temperature around 1000 °C and high magnetic moment up to 6.5 μ$_B$/f.u [9], both of which are important for device applications. Since the last decades, more and more Heusler alloys have been explored with the intention of searching of the property of half-metallicity. Among them, $Fe_2$-based Heusler alloys such as $Fe_2CoZ$ (Z = Ge, Ga) [10], $Fe_2CrZ$ (Z = Bi, Sb, As, P) [11], $Fe_2YP$ (Y = Zr, Mn, and Ti) [12-14], $Fe_2YAl$ (Y = Sc, Cr) [15], $Fe_2YAl$ (Y = Ni, Mn, Cr) [16], $Fe_2YGa$ (Y = Mn, Cr, V) [17], $Fe_2YSi$ (Y = Co, Cr, Mn) [18-20], and $Fe_2TiZ$ (Z = Sb, Sn, In, As, Ge and Ga) [21], have been extensively investigated by means of first-principles calculations. It is progressively being known that the physical properties of HAs are dependent on their structural order [22]. The site preference rule (SPR) has been widely used in the theoretical design of the full-Heusler (of type $X_2YZ$) alloys. This rule sets a demarcation between the formation of the $L2_1$ and XA phases of the full Heusler alloys. The detail description of the SPR will be presented in the results and discussions section of this paper. Some of the XA type full-Heusler alloys such as $Mn_2CoAl$ [23], $Ti_2MnAl$ [24], and $Ti_2CoSi$ [25] have been predicted to be a spin-gapless semiconductors using SPR. On the other hand, some of them e.g. $X_2CuAl$ (X= V, Ti, Cr, Sc, Mn, Zr, Hf), are found to be a perfect half-metals [26]. Interestingly, some counter-examples such as $Ti_2FeZ$ (Z = Ga, Al) [27], and our recent work on $Co_2FeAl$ Heusler alloy [28] don't follow the site preference rule (SPR). Moreover, F. Dahmane et al. [29] studied the $L2_1$ ($Cu_2MnAl$ prototype) and XA ($Hg_2CuTi$ prototype) ordering effect on the phase stability of $Fe_2XAl$ (X= Cr, Mn, Ni) Heusler compounds. They found that $L2_1$ phase of $Fe_2CrAl$ and $Fe_2MnAl$ are more stable as compared with XA phase at the equilibirium volume. Xiaotian Wang et al. [30], studied the $L2_1$ and XA ordering effect on Hafnium-Based full-Heusler Alloys and found that all of them were likely to exhibit the $L2_1$-type structure instead of the XA one. We also reported $L2_1$ and XA ordering effects on $Co_2FeAl$ [28], and found that XA (or indirect) ordered structure is much more stable than the $L2_1$ (direct) structure. Further, we have shown that the physical properties were not only dependent on $L2_1$ and XA ordering of atoms at their Wyckoff sites, but also strongly dependent on the exchange correlation potential (U).

Lalrinkima Siakeng et al. recently reported that $Fe_2CoAl$ alloy in $C1_b$ (or XA) phase [31], while the calculated electronic structures were not able to address the site preferences of atoms in their respective structures under different symmetries viz., $L2_1$ (space group (SG): $Fm\bar{3}m$) and XA (SG: $F\bar{4}3m$). Additionally, which magnetic state ( paramagnetic, ferromagnetic or antiferromagnetic) will be the ground state of FCA need to be explored to understand the complete electronic behavor of $Fe_2CoAl$ alloy.

The present systematic study addresses some important features, which may hinder the device compatibility and performance:

(i) Comprehensive reports are available for $Co_2FeAl$ (CFA) full Heusler alloys (see ref. [28] ,and references therein) where all literature says that CFA alloy tends to form $L2_1$ structure; however, our recent comparative study of $L2_1$ and XA ordering effect on phase stability of CFA reveals that XA phase is a ground state of CFA, not the $L2_1$. A very few reports are available on $Fe_2CoAl$ alloy and all of them are based on the XA (indirect/inverse) phase of FCA. Hence, we systematically studied the phase stability of FCA alloy under different cubic symmetries: $L2_1$ (SG: $Fm\bar{3}m$) and XA (SG: $F\bar{4}3m$). Additionally, we consider the stability of magnetic states, which are not addressed before.

(ii) Since decades, Slater Pauling rule is commonly used to predict the half metallic ferromagnets. Here we explore whether it has one to one relationship with the electronic properties of FCA. We also examine whether the site preference rule (SPR) is also suitable for $Fe_2CoAl$ alloy.

(iii) It is widely known that only generalised gradient approximation (GGA) is not sufficient to describe the complete electronic behavior of the systems having 3d electrons therefore, we also focus on the exchange and correlation (U) effects on electronic and magnetic properties of FCA.

Our results reveal that the physical properties are highly dependent on $L2_1$ and XA ordering of atoms at their Wyckoff sites. Furthermore, electronic structure is significantly affected in the presence of Coloumb potential (U). We found that XA-I type FCA alloy is energetically most stable and exhibit ~78% spin polarization at the Fermi energy ($E_F$) as compared with other structures. The site preference rule is also valid in $Fe_2CoAl$ alloy. Hence it is proficient for spintronics application. Present theoretical study will provide guidance for synthesis of the effcient $Fe_2CoAl$ based alloy.

## 2. Computational methods

All calculations were performed using Wien2k code [32] based on the full potential linearized augmented plane wave method. [33] Band-gap underestimation was known to occur with the consideration of the semi-local exchange-correlation function, and hence a generalized gradient approximation (GGA) of Perdew-Burke-Ernzerhof (PBE) was employed in the calculations to overcome it [34]. The effect of on-site Coulomb interaction (U) was also included in our calculations. The maximum l value ($l_{max}$) for the expansion of the wave function in spherical harmonics inside the atomic sphere was restricted to $l_{max} = 10$. The wave function in the interstitial region was expanded in plane waves with a cutoff of $R_{MT}K_{max} = 7$, where $R_{MT}$ represents the atomic radii of the smallest sphere and $K_{max}$ is the largest k vector. The charge density in the interstitial region was expanded up to $G_{max} = 12$ a.u$^{-1}$. A grid of 15×15×15 k-points was taken for the computations. The electronic and magnetic properties were studied at optimized lattice constants.

## 3. Results and discussions

3.1. $L2_1$ and XA ordering competition in $Fe_2CoAl$ Heuler alloy

The ground state of $Fe_2CoAl$ Heusler alloy is found, performing the lattice optimization calculation for different atomic occupations (as listed in Table. 1) in $L2_1$ and XA phases with various magnetic configurations viz., non-magnetic or paramagnetic (NM/PM), ferromagnetic (FM) and antiferromagnetic (AFM). The $X_2YZ$ type Heusler compounds such as $Fe_2CoAl$ may crystallize in conventional $L2_1$ ($Cu_2MnAl$) and inverse XA ($Hg_2CuTi$) structures under space group of $Fm\bar{3}m$ (space group number 225), and $F\bar{4}3m$ (space group number 216), respectively. A detailed description/definition of the difference in L2$_1$ and XA structures can be found in our previous paper [28]. There are four available Wyckoff sites: A (0, 0, 0), B (0.25, 0.25, 0.25), C (0.5, 0.5, 0.5), and D (0.75, 0.75, 0.75) along the body diagonal as shown in Fig. 1(a, b). In L2$_1$ (or direct) structure, Fe atoms occupy B (0.25, 0.25, 0.25) and D (0.75, 0.75, 0.75) sites, and Co/Al atoms occupy A (0, 0, 0) and C (0.5, 0.5, 0.5) sites, respectively. On the other hand, in XA (or inverse) structure, two inequivalent atoms of Fe (as labelled Fe1 and Fe2) occupy A (0, 0, 0) and D (0.75, 0.75, 0.75) sites and Co and Al atoms occupy C (0.5, 0.5, 0.5) and B (0.25, 0.25, 0.25) sites, respectively (see Fig.1). Generally, X$_2$YZ type of compounds prefer XA (Hg$_2$CuTi) prototype structure if Y atom has larger atomic number than the X (from the same period) atom [35].

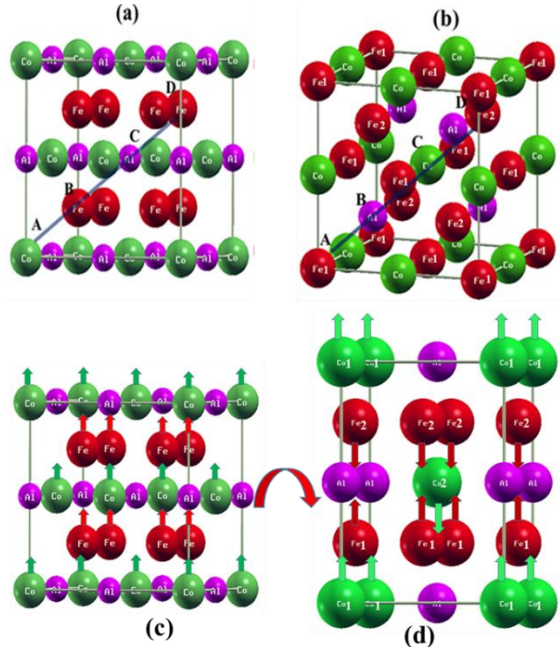

**Fig. 1.** Crystal structure of Fe$_2$CoAl (FCA) alloy in (a) L2$_1$ (Cu$_2$MnAl-prototype), and (b) XA (Hg$_2$CuTi prototype); (c, d) shows the arrangement of Co/Fe magnetic moments for ferromagnetic (FM) and antiferromagnetic (AFM) configuration, respectively. All crystal structures are generated using XCrysDen software [36].

The geometrical frustration is always present in the FCC lattice. To achieve antiferromagnetic (AFM) ordering, structural distortion is required and it can be one of the three types: (i) distortion is along the [001] direction (tetragonal distortion) that is seen in our case; (ii) distortion is along the [011] direction (orthorhombic distortion) and (iii) distortion is along [111] direction (rhombohedral distortion). The tetragonal distortion is usually observed in Heusler alloys [37]. A ferromagnetic arrangement is shown in Fig. 1(c). To obtain an AFM structure, we construct a supercell of FCA which takes the space group Pmmm (space group number 47) as shown in Fig. 1(d). Here, the ferromagnetic planes of Co spins (up/down) are alternatively arranged in a specific direction [001]. Similarly, ferromagnetic planes of Fe spins (up/down) are also arranged [38]. The non-magnetic (or paramagnetic) state means that all the constituents atoms of $Fe_2CoAl$ alloy have zero spin polarization (P) at Fermi energy ($E_F$).

The total energies ($E_{tot}$) have been calculated for both $L2_1$ and XA phases for the different volumes of the cell and the energy difference ($E_{tot}$-$E_0$) curves as a function of the unit cell volume are shown in Fig. 2.

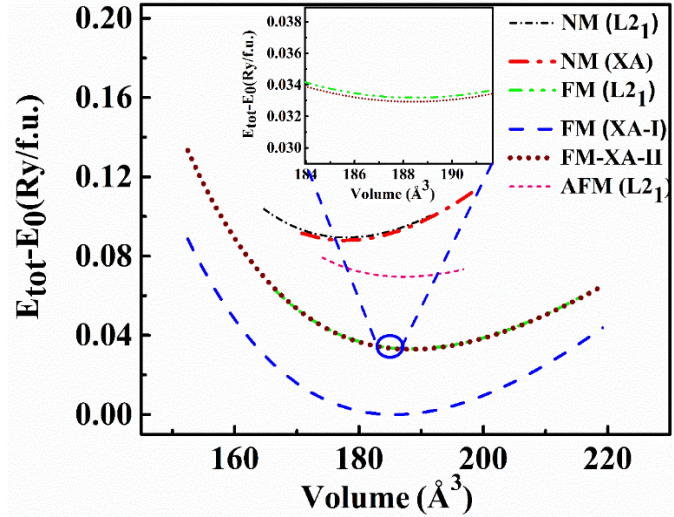

**Fig. 2.** (Color online) Total energy difference ($E_{tot}$-$E_0$) as a function of the unit cell volume of $Fe_2CoAl$ alloy under different possible nonmagnetic (NM) and magnetic configurations (ferromagnetic (FM) and antiferromagnetic (AFM)). Inset presents the magnified region shown by a blue circle.

It is noted that the total energy difference versus volume [($E_{tot}$-$E_0$) versus V] curves are obtained after fitting with Birch-Murnagham equation [39] and the fitted ground state parameters such as optimized lattice constants "$a_0$", equilibrium volume $V_0$, bulk modulus B and its pressure derivative B´ are shown in Table. 2. From Fig. 2, it is seen that $Fe_2CoAl$ alloy energetically prefers to crystallize in XA phase, but not in $L2_1$ phase. More specifically, it crystallizes in XA-I phase with ferromagnetic ordering as expected from the site preferences rule (SPR) [18, 40, 41]. The results reveal that the element with higher number of valence electrons prefers C (0.5, 0.5, 0.5) site and the element with lower number of valence electrons tends to enter at A (0, 0, 0) and D (0.75, 0.75, 0.75) Wyckoff sites, whereas the main group element Al usually prefers B (0.25, 0.25, 0.25) site. Hence, XA-I structure follows the site preference rule. A similar theoretical study was carried out on $Fe_2CoGa$ alloy, which has the same valence as $Fe_2CoAl$, and it was predicted that the inverse XA ($Hg_2CuTi$ prototype) structure was preferable [42]. The similar results also reported by our group elsewhere [28]. However, these results are in contrary with other's reports [30], where they predicted that the $L2_1$ phase is much more stable than the XA one. Therefore a comparative study of phase stability is crucial and should be addressed properly.

From data of Table. 2, it is clear that the optimized value of lattice constant "$a_0$" of XA-I (Hg$_2$CuTi prototype) structure is found to be 5.70 Å, which is much closer to the experimental one compared to the other structures. In the next sections, we shall focus in-depth studies on electronic structures of ferromagnetic XA-I (most stable) structure.

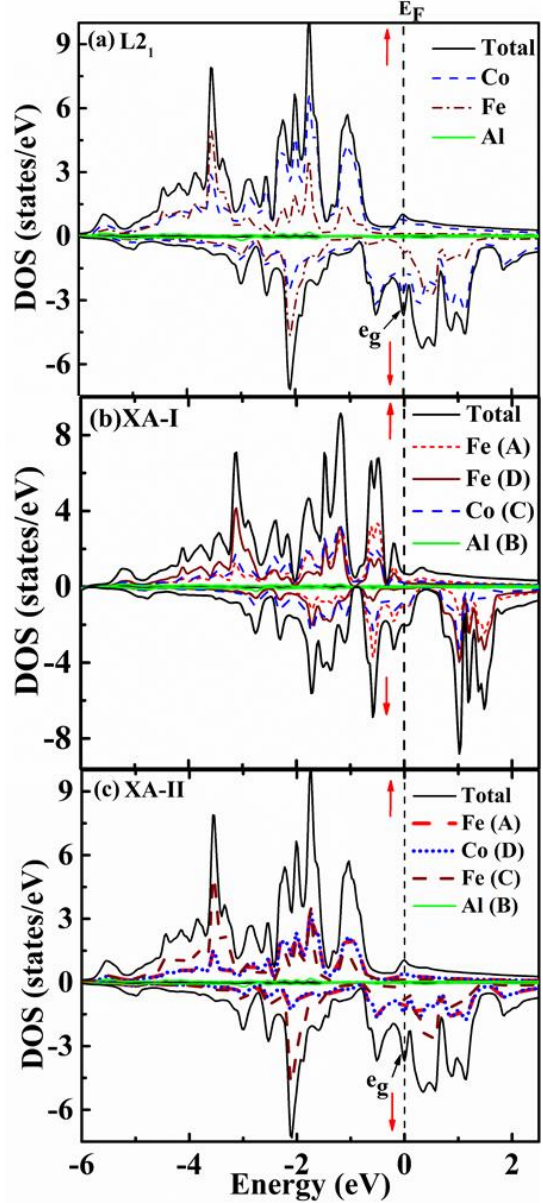

**Fig. 3.** Plots of total and element specific density of states (DOS) of Fe$_2$CoAl (FCA) in all atomic ordering of L2$_1$, XA-I, and XA-II types under (a) Cu$_2$MnAl and (b, c) Hg$_2$CuTi prototypes. Upper and lower panels represent DOS for up spin and down spin respectively as marked by the red arrows. Fermi level (E$_F$) was set at zero energy.

## 3.2. Electronic properties of L2$_1$ and XA types Fe$_2$CoAl within GGA

It is progressively being known from the theoretical electronic structures [43-45] that the physical properties of the Heusler alloys are strongly dependent on the atomic occupancies at their available Wyckoff sites in the unit cell and hence need to be explored. Therefore, in this section we discuss about the electronic behavior and magnetism of the L2$_1$ and XA types FCA. The spin polarization P, at Fermi energy (E$_F$) is defined by the following eq. 1.

$$P = \frac{D\uparrow(E_F) - D\downarrow(E_F)}{D\uparrow(E_F) + D\downarrow(E_F)} \qquad (1)$$

Where D$\uparrow$ (E$_F$) and D$\downarrow$ (E$_F$) represent the density of states at Fermi energy (E$_F$) for up and down spin channels, respectively. It displays a finite value for ferromagnetic materials and vanishes for paramagnetic and antiferromagnetic materials below Curie temperature [18]. Half-metallic materials exhibit 100% spin polarization only when D$\uparrow$ (E$_F$) or D$\downarrow$ (E$_F$) equals to zero at E$_F$. The calculated spin polarization values (P) for all type of ordered structures are shown in Table 3. We have calculated the total and atomic specific spin-polarized density of states (DOS) per electron volt (eV) at their respective optimized lattice parameters "a$_0$" under GGA approximation and shown in Fig. 3. For L2$_1$, XA-I and XA-II types Fe$_2$CoAl, their density of states (DOS) exhibit a common ferromagnetic metallic nature at Fermi energy. One can clearly see that the electronic structures of them remain unchanged whenever Fe/Co ordering is very different. The higher density peak at E$_F$ is mainly due to the Co/Fe e$_g$ state, which can be easily seen in Fig. 3(a, c). These peaks are responsible for the instability of regular L2$_1$ and XA-II phases compared with the XA-I phase. In the case of XA-I, this peak is totally shifted to the valance region and hence attributed to energetically most stable XA-I structure which was also confirmed from the volume optimization curve (see Fig. 2). The site preferences of atoms under L2$_1$ and XA-II type structures do not affect the general shape of the total density of states excluding XA-I. Moreover, in XA-I, double pseudogap structures (in the spin down channel) are observed just below (at about -0.8 eV) and above (at about 0.4 eV) the Fermi level (E$_F$), which imply the covalent bonding between the atoms [46]. Similar pseudogap structures have also been reported for inverse XA structure: Fe$_2$CoGa and Fe$_2$CoZn [42]. Half-metallicity is not observed in any possible structures as Fermi energy (E$_F$) has totally been shifted to the valance region and hence the spin polarization is reduced [47]. This argument is further supported by the calculated spin polarized band structures of the L2$_1$ and XA types FCA at their equilibrium lattice constants (a$_0$) as shown in the Fig. 4(a, b, c). The majority and minority bands of L2$_1$ and XA-II looks similar and are in close agreement with the DOS plots shown in Fig. 3(a, c). The band structure of XA-I, is highly affected by the influence of the atomic occupancies in the cell and that is why it shows quite different band structure than those of the L2$_1$ and XA-II. Although it exhibits a metallic nature since spin up and spin down bands are crossing at the Fermi level (E$_F$), however, a gap-like feature can be seen in the minority bands at X point.

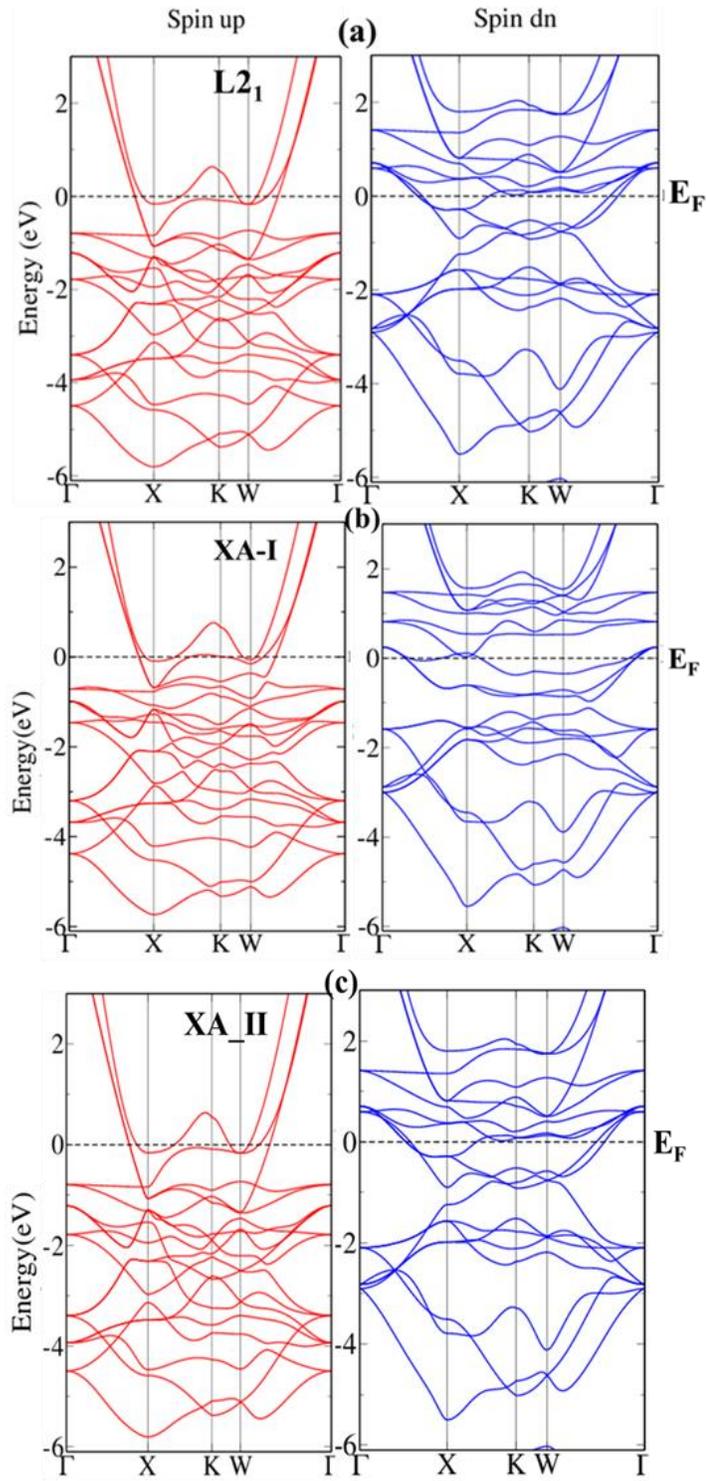

**Fig. 4.** Sown are the spin polarized band-structures of $Fe_2CoAl$ alloy in (a) $L2_1$, (b) XA-I and (c) XA-II phases. Fermi level $E_F$ has been shown by a dotted lines, red and blue curves show the spin up and spin down bands within GGA, respectively.

3.3 Magnetic Properties of L2$_1$ and XA types Fe$_2$CoAl

Wurmehl et al. [48] studied many Co base full Heusler-alloys and made a linear relationship between the total magnetic moment per unit cell M$_t$ (μ$_B$/f.u.) and Curie temperature, T$_c$ (K) as eq. 2.

$$T_c(K) = 23 + 181 \times M_t\ (\mu_B/f.u.) \qquad (2)$$

This equation gives the highest T$_c$ for Half-metallic materials, which have a large magnetic moment. Using this equation, T$_c$ of Co$_2$FeSi has been reported to be higher than 1000 K, which was consistent well with experimental results [48]. A linear variation of T$_c$ with magnetic moment M$_{tot}$ was also reported by Kubler et al. [49] on the following compounds e.g., Co$_2$TiAl, Co$_2$VGa, Co$_2$VSn, Co$_2$CrGa, Co$_2$CrAl, Co$_2$MnAl, Co$_2$MnSn, Co$_2$MnSi, and Co$_2$FeSi. The calculated Curie temperatures using eq. 2 [48, 50] are tabulated in Table. 3 along with experimental results for comparison. These results are consistent with recent DFT study too reported by Lalrinkima Siakeng et al. [51]. From Table. 3, it is clear that the net magnetic moment is contributed due to Fe and Co atoms and the contribution from Al is insignificant in all structures. Moreover, one can also see that the spin polarizations and Curie temperatures are high to those which have higher magnetic moments i.e. L2$_1$ and XA-II structures than the XA-I structure. Any structures don't exhibit the integer magnetic moments and far away from the Slater-Pauling value [52], which is 4.0 μ$_B$ in this case. In a search of half-metallic materials, we conclude that these two conditions will still remain necessary to meet the criteria of exhibiting half-metallicity. The total and partial magnetic moments including spin polarization values and Curie temperatures are nearly equal for all structures excluding XA-I type. The experimental value of the total magnetic moment per cell, which is 4.9 μ$_B$/f.u. (see Table. 3), is in close agreement with the XA-I (most stable) structure.

3.4 Effect of on-site Coulomb interaction (U)

Here the results of the electronic structure calculation of most stable XA-I phase of Fe$_2$CoAl alloy using GGA+U (Coulomb potential) are presented. Noted that we have treated all possible structures under GGA+U, but did not get significant changes in the density of states (other data are not shown here). Typically, Heusler alloys are endorsed to exhibit localized moments; so electron correlation can play an important role [48]. The GGA+U scheme is used to calculate the electronic structure to understand whether the addition of correlation resolves the inconsistency between the theoretical and experimental magnetic moment. In Wien2k, the effective Coulomb-exchange interaction, $U_{eff} = U - J$, where U and J are the Coulomb and exchange parameter, is used to account for double counting corrections. In Fig. 5., the spin-polarized total and partial density of states (DOS) are shown using the GGA+U method. The effective Coulomb-exchange parameters were set to $U_{eff.Co}$ = 4.22 eV and $U_{eff.Fe}$ = 4.35 eV [53] at the Co and Fe sites, respectively. As seen from Fig. 6, a small energy gap of 0.87 eV is observed in minority spin channel at about 0.5 eV above the $E_F$. The Fermi energy ($E_F$) cuts the minority bands below the gap is due to the region that the total magnetic moment per cell is too high than the predicted Slater-Pauling value and is not an integer, which is expected for a half metal [48]. Additionally, a similar kind of gap-like structure has also been reported previously in Refs [31, 48, 54].

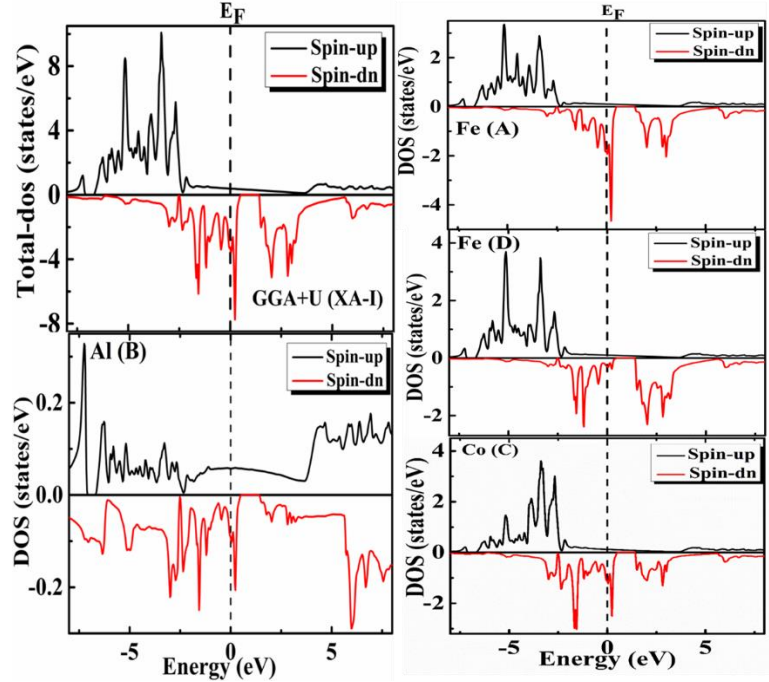

**Fig. 5.** Plots of spin-polarized total and atomic-resolved density of states of $Fe_2CoAl$ (FCA) under GGA+U approximation.

Hence, FCA alloy is still not a half-metal; however, spin polarization is as high as 78% which is very important for spintronics application. These values were underestimated when dealt with GGA approximation. The metallic nature of FCA alloy is further confirmed from the results of the spin polarized band structures calculation shown in Fig. 6. To achieve a half-metal, Fermi level could be tuned in the middle of the gap by varying the lattice constants "a". [31, 48] To explain the origin of the energy gap in the down spin channel, atomic resolved DOS has also been presented in Fig. 5. We observed that the bonding and antibonding states are formed due to the d-d hybridization between the low valence Fe (A)/Fe (D) atoms having d states in higher energy, and high valence Co (C) atoms with d states in lower energy; hence attributed to the gap-like feature in the spin down channel.

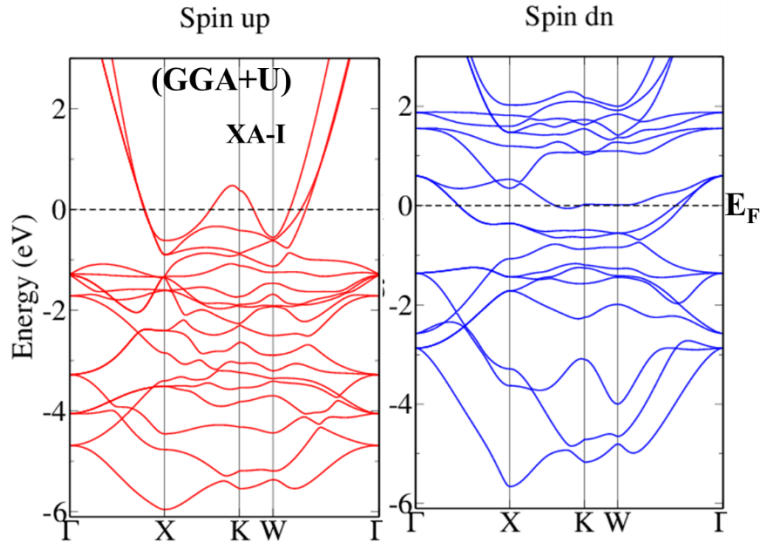

**Fig. 6.** Spin polarized band-structures of XA-I type $Fe_2CoAl$. Fermi level $E_F$ has been shown by a dotted lines; red and blue curves show the spin up and spin down bands within the GGA+U.

## 4. Conclusion

The effects of atomic ordering in their Wyckoff sites on phase stability, electronic structures and magnetic properties of conventional $L2_1$ and inverse XA type $Fe_2CoAl$ alloy have been investigated and compared. FCA alloy energetically favored the inverse XA-I structure compared with the others structures, as expected from the site preference rule. Electronic, structural and magnetic properties were not much affected by atomic site preferences in ferromagnetic $L2_1$ and XA-II structures but significant changes have been observed in the case of XA-I structure. Moreover, only this structure has shown the possibility to achieve a half-metal because of having a finite spin down energy gap within GGA+U. Atomic site preferences did not restore the half-metallicity in any structures of the $Fe_2CoAl$ (FCA) alloy but the variation of lattice constants worked well. FCA alloy is a ferromagnetic (FM) metal under all kind of possible structures and is found to be most stable in XA-I structure with a spin polarization value of around ~78% at $E_F$. Hence, we conclude that FM-FCA alloy in XA-I ordering could be used as a highly spin-polarized material in spintronics devices.

## 5. Acknowledgments

Aquil Ahmad sincerely acknowledges the University Grant Commission (UGC) Delhi, MHRD Delhi, India for providing fellowship for Ph.D. work. A. K. Das acknowledges the financial support of DST, India (project no. EMR/2014/001026). Our fruitful discussion with Dr. Monodeep Chakraborty is highly acknowledged. We also acknowledge our departmental computational facility, IIT Kharagpur, India.

**Author contribution statement**

Aquil Ahmad performed all the calculations and manuscript writing under the supervision of Dr. A.K. Das and Dr. S.K. Srivastava.

**REFERENCES**

[1] A. Fert, Thin Solid Films, **517** 2 (2008).
[2] S. Wurmehl, G.H. Fecher, H.C. Kandpal, V. Ksenofontov, C. Felser, H.J. Lin, Appl. Phys. Lett., **88**, 032503 (2006).
[3] R.A. de Groot, F.M. Mueller, P.G.v. Engen, K.H.J. Buschow, Phys. Rev. Lett., **50**, 2024 (1983).
[4] J.M. Coey, M. Venkatesan, C.B. Fitzgerald, Nat. mater., **4**, 173 (2005).
[5] W.Z. Xiao, L.l. Wang, B. Meng, G. Xiao, RSC Adv., **4**, 39860 (2014).
[6] A. Ahmad, S.K. Srivastava, A.K. Das, arXiv:1906.06516, (2019).
[7] Y. Miura, S. Muramoto, K. Abe, M. Shirai, Phys. Rev. B, **86**, 024426 (2012).
[8] I. Galanakis, P. Mavropoulos, J. phys. Cond. Matter, **19** 315213 (2007).
[9] A. Ahmad, S. Mitra, S. Srivastava, A. Das, J. Magn. Magn. Mater., **474,** 599 (2019).
[10] K. Seema, M. Kaur, R. Kumar, J. Integ. Sci. Tech., **1**, 41 (2013).
[11] B.S. Chen, Y.Z. Li, X.Y. Guan, C. Wang, C.X. Wang, Z.Y. Gao, J. Superc. Novel Magn., **28**, 1559 (2014).
[12] N. Kervan, S. Kervan, Intermetallics, **37,** 88 (2013).
[13] N. Kervan, S. Kervan, Intermetallics, **24**, 56 (2012).
[14] S. Babiker, G. Gao, K.L. Yao, J. Superc. Novel Magn, **31**, 905 (2017).
[15] N. Arıkan, A. İyigör, A. Candan, Ş. Uğur, Z. Charifi, H. Baaziz, G. Uğur, J. Mater. Sci., **49**, 4180 (2014).
[16] F. Dahmane, Y. Mogulkoc, B. Doumi, A. Tadjer, R. Khenata, S. Bin Omran, D.P. Rai, G. Murtaza, D. Varshney, J. Magn. Magn. Mater., **407**, 167 (2016).
[17] L.X. H.L. Jing-Bo, L.Y. H.J. Y. Jiu, Acta Physica Sinica, 10, 085 (2011).
[18] H.C. Kandpal, G.H. Fecher, C. Felser, J. Phys. D Appl. Phys., 40, 1507 (2007).
[19] H. Luo, Z. Zhu, L. Ma, S. Xu, H. Liu, J. Qu, Y. Li, G. Wu, , J. Phys. D Appl. Phys., **40**, 7127 (2007).
[20] Y. Du, G.Z. Xu, X.M. Zhang, Z.Y. Liu, S.Y. Yu, E.K. Liu, W.H. Wang, G.H. Wu, Crossover of magnetoresistance in the zero-gap half-metallic Heusler alloy $Fe_2CoSi$, EPL **103**, 37011 (2013).
[21] H. Luo, G. Liu, F. Meng, J. Li, E. Liu, G. Wu, J. Magn. Magn. Mater., **324**, 3299 (2012).
[22] X. Wang, Z. Cheng, H. Yuan, R. Khenata, J. Mater. Chem. C, 5, 11559 (2017).
[23] I. Galanakis, K. Özdoğan, E. Şaşıoğlu, S. Blügel, J. appl. phys., 115, 093908 (2014).
[24] W. Feng, X. Fu, C. Wan, Z. Yuan, X. Han, N.V. Quang, S. Cho, phys. status solidi Rapid Res. Lett., **9**, 641 (2015).
[25] A. Birsan, P. Palade, V. Kuncser, J. Magn. Magn. Mater., **331**, 109 (2013).
[26] H. Luo, Y. Xin, B. Liu, F. Meng, H. Liu, E. Liu, G. Wu, J. Alloys Comp., **665**, 180 (2016).
[27] X. Zhang, Z. Liu, Y. Zhang, H. Liu, G. Liu, Y. Cui, X. Ma, Intermetallics, **73**, 26 (2016).
[28] Aquil Ahmad, S.K. Srivastava, A.K. Das, J. Magn. Magn. Mater., 491, 165635 (2019).
[29] F. Dahmane, Y. Mogulkoc, B. Doumi, A. Tadjer, R. Khenata, S.B. Omran, D. Rai, G. Murtaza, D. Varshney, J. Magn. Magn. Mater., **407**, 167 (2016).
[30] X. Wang, Z. Cheng, W. Wang, Materials, **10**, 1200 (2017).

[31] L. Siakeng, G.M. Mikhailov, D. Rai, J. Mater. Chem. C, 6, 10341 (2018).
[32] P. Blaha, K. Schwarz, G. Madsen, D. Kvasnicka, J. Luitz, WIEN2K, ISBN 3-9501031-1-2, 2002.
[33] D. Singh, Plane waves, London: Kluwer Academic Publishers, 1994.
[34] J.P. Perdew, K. Burke, M. Ernzerhof, Phys. Rev. lett., **77**, 3865 (1996).
[35] N. Xing, Y. Gong, W. Zhang, J. Dong, H. Li, Comput. Mater. Scie., **45**, 489 (2009).
[36] A. Kokalj, Computational Materials Science, **28**, 155 (2003).
[37] Y.I. Matsushita, G. Madjarova, J.K. Dewhurst, S. Shallcross, C. Felser, S. Sharma, E.K. J. Phys. D Appl. Phys., 50, 095002 (2017).
[38] Z. Aarizou, S. Bahlouli, M. Elchikh, Modern Phys. Lett. B, **29**, 1550093 (2015).
[39] F. Birch, Phys. Rev., 71, 809 (1947).
[40] H. Luo, Z. Zhu, L. Ma, S. Xu, X. Zhu, C. Jiang, H. Xu, G. Wu, J. Phys. D Appl. Phys., 41, 055010 (2008).
[41] Y. Han, Z. Chen, M. Kuang, Z. Liu, X. Wang, X. Wang, Results Phys., **12**, 435 (2019).
[42] A. Dannenberg, M. Siewert, M.E. Gruner, M. Wuttig, P. Entel, Phys. Rev. B, **82** (2010).
[43] L. Bainsla, A. Mallick, M.M. Raja, A. Coelho, A. Nigam, D.D. Johnson, A. Alam, K. Suresh, Phys. Rev. B, **92**, 045201 (2015).
[44] X. Wang, Z. Cheng, R. Khenata, Y. Wu, L. Wang, G. Liu, J. Magn. Magn. Mater., **444**, 313 (2017).
[45] F. Ahmadian, J. superc. novel magn., **26**, 381 (2013).
[46] C.M. Li, H.-B. Luo, Q.-M. Hu, R. Yang, B. Johansson, L. Vitos, Phys. Rev. B, **82**, 024201 (2010).
[47] I. Galanakis, P. Mavropoulos, P.H. Dederichs, J. Phys. D Appl. Phys., 39, 765 (2006).
[48] S. Wurmehl, G.H. Fecher, H.C. Kandpal, V. Ksenofontov, C. Felser, H.J. Lin, J. Morais, Phys. Rev. B, 72, 184434 (2005).
[49] J. Kübler, G.H. Fecher, C. Felser, Phys. Rev. B, **76**, 024414 (2007).
[50] X.Q. Chen, R. Podloucky, P. Rogl, J. Appl. Phys., 100, 113901 (2006).
[51] L. Siakeng, G.M. Mikhailov, D.P. Rai, J. Mater. Chem. C, **6**, 10341 (2018).
[52] I. Galanakis, E. Şaşıoğlu, S. Blügel, K. Özdoğan, Phys. Rev. B, **90** (2014).
[53] D. Rai, A. Shankar, J. Sandeep, L. Singh, M. Jamal, S. Hashemifar, M. Ghimire, R. Thapa, Armenian J. Phys., **5,** 105 (2012).
[54] P. Entel, M.E. Gruner, A. Dannenberg, M. Siewert, S.K. Nayak, H.C. Herper, V.D. Buchelnikov, Materials Science Forum, **635**, 3 (2009).

# Supplementary Data for

## Competition of L2$_1$ and XA Ordering in Fe$_2$CoAl Heusler Alloy: A First-Principles Study

Aquil Ahmad[*], A. K. Das, and S. K. Srivastava

*Department of Physics, Indian Institute of Technology Kharagpur, India-721302*

[*]*Email:* aquil@phy.iitkgp.ernet.in

| Structure | Wyckoff sites | | | |
|---|---|---|---|---|
| | **A** | **B** | **C** | **D** |
| L2$_1$ | Co | Fe | Al | Fe |
| XA-I | Fe1 | Al | Co | Fe2 |
| XA-II | Fe1 | Al | Fe2 | Co |

**Table S1:** Structural order of regular L2$_1$ (Cu$_2$MnAl) and inverse XA (Hg$_2$CuTi) prototype Heusler alloys, where A, B, C and D represent the available Wyckoff sites: (0, 0, 0), (0.25, 0.25, 0.25), (0.5, 0.5, 0.5) and (0.75, 0.75, 0.75), respectively along the body diagonal.

| Parameter | L2$_1$ (Cu$_2$MnAl prototype) | XA (Hg$_2$CuTi - prototype) | |
|---|---|---|---|
| | | XA-I | XA-II |
| Exp. lattice constants (Å) | (5.71 [1]) | | |
| Equilibrium lattice constants, a$_0$ (Å) | 5.73 | 5.70 (5.701)[a] | 5.73 |
| Bulk modulus B (GPa) | 154.98 | 175.70 | 154.68 |
| Derivative of Bulk modulus (B´) | 5.48 | 4.99 | 5.68 |
| Equilibrium volume, V$_0$ (Å$^3$) | 188.43 | 185.66 | 188.47 |

[a] Ref. [2]

**Table S2:** The calculated optimized lattice parameter, a$_0$ (Å), equilibrium volume V$_0$ (Å$^3$), bulk modulus B (GPa) and its pressure derivative B´ of the Fe$_2$CoAl alloy in Cu$_2$MnAl and Hg$_2$CuTi structure under different atomic arrangements to their respective Wyckoff sites. The other theoretical and experimental results are shown in parenthesis for comparison.

| Structure | $M_{Fe}$ (μB) | $M_{Co}$ (μB) | $M_{Al}$ (μB) | $M_{Tot}^{Cal}$ (μB) | P% | $M_{Tot}^{Rep}$ (μB) | $M_{Sp}$ (μB) | $T_C^{Cal}$(K) | $T_C^{Exp}$ (K) |
|---|---|---|---|---|---|---|---|---|---|
| L2$_1$ | 2.04 | 1.85 | -0.05 | 5.67 | 57 | 5.05[a] [3] | 4.0 | 1049 | 1130[5], 1010 - 1073[4] |
| XA-I | Fe1/Fe2 = 1.55/ 2.53 | 1.09 | -0.06 | 4.90 | 47 | 5.14[b] [1] | | 911 | |
| XA-II | Fe1/Fe2 = 2.03/ 2.03 | 1.85 | -0.53 | 5.66 | 55 | 4.9[c] [4] | | 1047 | |

[a, b] Theoretical values of FCA in XA phase (at T=0K)

[c] Experimental value of A2-disordered FCA under regular L21 structure (at T=300 K)

**Table S3:** The calculated total magnetic moments ($M_{Tot}^{Cal}$), and partial magnetic moments ($M_{Fe}$, $M_{Co}$, and $M_{Al}$) in μB, the spin polarization P (%), reported magnetic moments ($M_{Tot}^{Rep}$), Slater Pauling value ($M_{Sp}$), calculated and experimental Curie temperatures ($T_C^{Cal}$, $T_C^{Exp}$) are listed along with those of other structures. The results from other theoretical and experimental works are also given in brackets.

REFERENCES

[1] Matsushita Y.I. et al., J. Phys. D Appl. Phys., **50** (2017) 095002.
[2] Gilleben M. et al., Journal of computational chemistry, **31** (2010) 612-619.
[3] Siakeng L., Mikhailov G.M., Raid., J.Mater. Chem. C, **6** (2018) pp. 10341-10349
[4] Jain V. et al., AIP Confer. Proceed. **1536**, (2013) pp. 935-936.
[5] Hasier J., EPJ Techniques and Instrumentation, **4** (2017) 5.